# White holes, primordial black holes and Dark Matter

Bikash Sinha
Variable Energy Cyclotron Centre
1/AF, Bidhan Nagar, Kolkata – 700064, India
bikash@vecc.gov.in; bsinha1945@gmail.com

## Abstract

Strange Quark Nuggets are the relics of the microsecond old universe after the big bang. The universe having experienced a mini inflation of 7-e folding has gone through supercooling leading to a first order phase transition from quark to hadrons. Strange Quark Nuggets, the relics of this phase transition will constitute 12.5% or at the most 25% of the dark matter of the universe. Only $10^{-3}$ or $10^{-4}$ part of the total number of SQNs form binaries, and only some of the binaries will turn to black holes. Primordial Black Hole (PBH) as a relic is shown to be not a plausible scenario. Simultaneous observation of high energy gamma rays from SQN binaries formed from coalescence of SQNs along with black holes will be the vital cursor of this evolutionary scenario just presented.

## I. Introduction

It was recently observed by many authors [1, 2, 3, 4, 5, 6] that microseconds after the big bang the expanding universe experienced supercooling accompanied by mini inflation, of 7-e folding leading to a first order phase transition from quarks to hadrons.

The relics, in the form of quark nuggets with a baryon number larger than a critical number ~ $10^{43}$ will survive the evolution of the universe from that primordial epoch to now [6].

Rather more recently [8, 9], it was demonstrated that tunnelling of neutrons from the quark nuggets to the physical vacuum can be conjectured to be equivalent to Hawking radiation from a gravitational black hole [8, 9]. Thus, when the Hawking radiation gets turned off, the tunnelling of neutrons stop, and the nuggets survive the evolution of the universe. Since physical vacuum has no colour degrees of freedom the surviving quark nuggets are often referred to as white holes [8. 9], the colour remains confined inside the nugget.

It has been observed sometime ago [1] that these quark nuggets consist of strange quarks, the lowest energy state of the fermions, in the quark sector. Strange Quark Nuggets (SQNs) can be candidates of MACHO (Massive Astrophysical Cosmic Hals Objects) [10] observed by Alcock et. al. [11, 12, 13] sometime ago by microlensing towards Large Megallinic Cloud LMC.

In this work, I wish to investigate all possible sources of cosmic quark matter in the universe with the primary motivation of estimating how much of the dark matter is due to cosmic quark matter.

It is further noted that the mass of the binaries, created by the coalescence of SQNs as well as neutron stars may indeed go beyond the Chandrasekhar limit [14] and turn to black holes. These black holes will also be candidates of dark matter. The SQNs and their binaries in the universe, could clump under gravity and go on to form some sort of quark galaxy or at least mimic a quark galaxy, which may be called as MQG. Quark stars, may also evolve from such a galaxy although the present wisdom indicates hadron to quark phase transition in the core of the neutron star is the source of quark stars. Quark stars are also part of the dark matter.

So, what are the source of cosmic quarks [15]?

Quark nuggets, relics of the first order phase transition from quarks to hadrons, microseconds after the big bang is a possible source; the core of the neutron stars, most likely to be inhabited by quarks, ejected after collision of the neutron stars, not necessarily head on, an event rather rare, but from non central collision will contribute to the quarks of the universe.

Finally, quarks from the quarks stars [15] are also part of the dark matter. One of the feature of the these strange quark stars is that they are not product of stellar evolution. They are most likely to be produced due to hadron – quark phase transition inside the core of the neutron star. Clearly, the maximum mass of these stars would be of the order of the mass of the neutron stars and indistinguishable from neutron stars at least in terms of their cosmological bulk properties.

It has already been claimed that SQNs are MACHOs [10], although some claim [16] MACHOs are candidates of Primordial Black Holes (PBH). This dichotomy will be examined in the following.

With that introduction, I wish to compute in the following the radius of the binary (using the celebrated Hawking entropy relation) [9, 17, 18] formed from the individual SQNs.

I then compute the mass of the binary, (assuming its spherical for simplicity) and then attempt to examine the criterion for the binaries turning to black holes.

## II. White Holes, Black Holes and Dark Matter

The Strange Quark Nuggets (SQNs), just after formation are under two kinds of field of forces, the gravitational field due to other SQNs and the radiation field generated from the ambient universe. It has been shown by us [10] that below a critical temperature, gravitational force overcomes the radiation pressure and the SQNs would begin to coalesce under mutual gravity of the SQNs. The mass of the SQNs was estimated for the clumped SQNs [10]. The clumped SQNs would then manifest themselves as MACHOs [10], observed by Alcock et. al. [11, 12, 13] sometime ago.

What are the total number of $N_{macho}$, within the horizon today?

With the cosmic microwave background temperature of ~3°K and time ~ $4 \times 10^{17}$ secs since the big bang, and the photon to baryon ratio $\eta = \gamma/s \approx 10^{-10}$, the total number of baryons content of the Cold Dark Matter (CDM) is ($\Omega_{CDM}/\Omega_B$) times the total number of baryons. For $\Omega_{CDM} \sim 0.3$ and $\Omega_B \sim 0.01$, the total number of baryons comes out to be ~ $1.6 \times 10^{79}$. It is known [10] that the total number of baryon in a MACHO are $2.44 \times 10^{56}$ and $1.13 \times 10^{55}$ for the baryon content of the nugget being $10^{42}$ and $10^{44}$ respectively. Thus, dividing the total number of baryons in CDM by that in a MACHO, $N_{macho}$ comes out to be in the range $N_{macho} \sim (0.7 \times 10^{23}$ to $1.4 \times 10^{24})$.

The most probable mass of a MACHO is $0.5 \, M_{\odot}$ but safely in the range $(0.5 - 1.0)$ $M_{\odot}$. It is known that the total dark matter mass is $(4 \times 10^{23}) \, M_{\odot}$ in the universe.

We immediately see that for a mass of $0.5 \, M_{\odot}$ of a SQN with $N_{macho} \sim 10^{23}$, the contribution to the dark matter of the universe from SQNs is 12.5% and for a mass of $1 M_{\odot}$ of a SQN, the contribution is 25%. However for $N_{macho} \sim 10^{24}$, the mass contribution of SQNs to the dark matter exceeds the total dark matter content of the universe, an implausible scenario. So, from this rather simple argument, it is felt that $N_{macho}$ is of the order of ~ $10^{23}$ with a mass in the range $(0.5 - 1.0) \, M_{\odot}$. With such large numbers some degree of uncertainty is implicit. The numbers quoted here are the lower limit, they can be much higher even by a factor of two or three.

If on the other hand MACHO are candidates of Primordial Black Holes (PBH) as has been suggested, the entire dark matter of the universe should consist of PBH only, an extremely unlikely scenario. One or more dark galaxies consisting of PBH only, as the sole contributor of dark matter is not a very plausible scenario.

It is therefore most likely that SQNs contribute between 12.5% to 25% of the dark matter of the universe. The rest of the dark matter consists of other species, as per the contemporary wisdom [7].

Possible observables to find out more precisely the contribution of SQNs and its binaries and the Black Holes formed from some of the binaries are discussed in the next section.

It now emerges, that the total mass of the MACHO is ~ (0.5 – 1.0) x $10^{23}$ $M_\odot$ considerably larger than the mass of our milky way which is of the order of ~ (1.5 x $10^{12}$) $M_\odot$. As mentioned one has the likely scenario of some sort of Strange Quark Galaxy or at least a mimic of a Strange Quark Galaxy, which I have called MQG [20].

The quark stars can ofcourse evolve from the core of the neutron star as has been mentioned. Their mass will be of the same order as the neutron star mass.

With that preamble let me now compute the radius of the binary system of two SQNs, first by using Hawking's celebrated [17, 18] entropy relationship as well as using the ansatz used by the present author [9, 15]

Hawking's entropy – surface area relationship gives the following simple equation [9, 17, 18].

$$S_{binary} = S_1^{SQN} + S_2^{SQN}$$

$$\text{leading to,} \quad \frac{\pi R_T^2}{G} = \frac{\pi R_1^2}{G} + \frac{\pi R_2^2}{G} \qquad (1)$$

Where $R_T$ is the radius of the binary, after the two SQNs coalesce, $R_1$ and $R_2$ being the individual radius of the SQNs.

$$\text{Thus, } R_T = \sqrt{R_1^2 + R_2^2} \qquad (2)$$

It is understood that since the SQNs are isolated system, the entropy will be conserved.

For simplicity we assume $R_1 = R_2 = R_o$ so that

$$R_T = \sqrt{2} \ R_0 = 1.4 \ R_0 \qquad (3)$$

Using the heuristic ansatz suggested by the present author [9, 15]

$$S = \pi R_T^2/G = [\pi R_1^2 B^{1/2} \ (M_1/M_\odot)^{-2} + \pi \ R_2^2 B^{1/2} \ (M_2/M_\odot)^{-2}] \qquad (4)$$

$$G = \tfrac{1}{2}\sigma : \sigma = 0.16 \ GeV^2 : M_1/M_\odot = M_2/M_\odot \approx 0.27$$

Which immediately leads to

$$R_T \cong 1.2\ R_0 \qquad\qquad (5)$$

remarkably close to the value obtained previously using eqn (3) – both the methods however are primarily based an Hawking's entropy surface relation $S_H = A/4L_B^2$ where A is the surface area and $L_B$, the length scale [9, 15, 17].

The mass of the binary formed by coalescence of two SQNs or for that matter two neutron stars can be written out [14]

$$M_B = \int_0^{R_T} 4\pi^2 \rho(r) dr \qquad\qquad (6)$$

For simplicity I assume the density of the binary is approximately constant throughout the whole volume of the binary thus, one has [14]

$$M_B = \frac{4}{3}\pi R_T^2 B + e_F N \qquad\qquad (7)$$

$e_F$, being teh Fermi energy per particle [14]

$e_F = \frac{3}{4}(\frac{9\pi}{2g})^{1/3} \frac{N}{R_T}$ where B is the Bag constent, N, the total number of fermions in the binary of radius $R_T$, g being the statistical degeneracy factor; after some elementary algebra

$$M_B = \frac{16}{3}\pi\ BR_T^3 \qquad\qquad (8)$$

In Table I, the mass of the possible binary $M_B$ is shown with various value of $R_1 = R_2 = R_0$, $R_T$ value are taken from eqn (3) based on Hawking entropy relation.

**Table 1**               $R_T = 1.4\ R_0 : B^{1/4} = 145.0$ MeV

| $R_0 = 3$ kms | $R_T = 4.2$ kms | $M_B/M_\odot = 0.25$ |
|---|---|---|
| $R_0 = 5$ kms | $R_T = 7.00$ kms | $M_B/M_\odot = 0.8$ |
| $R_0 = 7$ kms | $R_T = 9.8$ kms | $M_B/M_\odot = 1.10$ |
| $R_0 = 9$ kms | $R_T = 12.6$ kms | $M_B/M_\odot = 1.6$ |

Please note that the first three cases, although binaries, are not black holes.

I now compare these results to the maximum mass possible for strange quark stars estimated both by Witten [1] and Banerjee et. al. [14]. Witten [1] computed the maximum value of the mass of the strange quark star by numerically integrating the standard equations of stellar structure $p = \frac{1}{3}(\rho - 4B)$ whereas Banerjee et.al. [14] computed the value of the maximum mass by using the simple energy balance relations used by Landau for white dwarfs and neutron stars. For the maximum mass the results are very similar, as is expected.

$$M_{max}^{w} = 0.0258/(G^{3/2}B^{1/2}) \qquad (9)$$

$M_B = \frac{16}{3}\pi BR_T^{3}$, as mentioned before

For $R_1 = R_2 = R_0$ and $R_T \approx 12.6$ kms, the mass of the binary comes out $M_B/M_\odot \sim 1.6$ as shown in the table and for $R_T \cong 12.0$ kms it is $M_B/M_\odot \sim 1.58$ using Witten's formulation [1]. Banerjee et. al. also obtains [10] $M_B/M_\odot \sim 1.54$ for $R_T = 12.11$ kms. All these numbers are fairly close to each other. The conclusion I draw from the above discussion is that for a mass of 1.6 $M_B/M_\odot$ [10] of the binary, the star already becomes gravitationally unstable and turns to a black hole, still a candidate of dark matter. The same conclusion in general holds for the neutron star binaries.

Uptil now, one has been referring to the universe, the next obvious question is how many MACHOs inhabit the Milky way? The total visible mass of the Milky way is $\sim (1.6 \times 10^{11})$ $M_\odot$ corresponding to $2 \times 10^{68}$ baryons. This corresponds to a factor $2 \times 10^{-9}$ of all the visible baryons within the present horizon. Now, scaling the clumped SQNs within the horizon by the same factor yields the total number of MACHO in the Milky way, $N_{macho} \sim 10^{13-14}$, for a baryon number of the initial nuggets of $10^{42-44}$.

We have enough input of results to go over to the next section of "Results and Observables" and finally end with "Outlook".

## III. Results and Observables

The Strange Quark Nuggets (SQNs) which survive from the primordial epoch of the universe microseconds after the big bang should have a baryon number $10^{43}$ or above [6]. It is understood that just prior to the phase transition from quarks to hadrons, the universe experienced a mini inflation of 7-e folding precipitating a first order phase transition. The SQNs can be identified with MACHOs [10]. There are also suggestion [16] that MACHOs could be Primordial Black Holes (PBH). We find this is an implausible scenario. One of the reasons is that the density contrast required for the formation of PBH has to be much larger than the existing density contrast in the inflationary universe a scenario adapted in this work as in other similar work [3, 4, 5]. Second, it is unlikely that the entire dark matter of the universe is just from black holes only, which is the inevitable conclusion if MACHOs are PBH.

It has been demonstrated by us that the total number of MACHOs in the universe is $N_{Macho} \sim 10^{23-24}$ of mass typically $(0.5 - 1.0)$ $M_\odot$ [10]

From now on I identify SQNs with the MACHO. It is known that the mass of the entire dark matter of the universe is $\sim (4 \times 10^{23})$ $M_\odot$; for a mass of the SQN ~ 0.5 $M_\odot$, SQNs constitute 12.5% of the dark matter mass and for a mass of 1 $M_\odot$ of the SQNs constitute ~ 25% of the dark matter. For such large numbers the uncertainty is implicit.

For $N_{macho} \cong 10^{24}$ the SQNs exceeds the total mass of dark matter an unacceptable scenario.

Two points emerge: the mass of the SQNs is most likely 0.5 $M_\odot$ as has been suggested from various view points of views [7], second MACHO being a candidate of PBH is unlikely,

My aim in this work is not so much on gravitational waves, specifically but on the quark matter contribution to the dark matter. I shall briefly in the following refer to gravitational waves in passing to seek out the observational window for the black holes originating from SQN binaries BHSQN. Not all SQNs turn to binaries or black holes, so, there exists a substantial mass of SQNs as dark matter over and above BHSQNs, also candidates of dark matter.

From Nakamura et. al. [19] it can be safely concluded that the number of binaries is approximately $(10^{-3}, 10^{-4})$ times the total number of SQNs. Out of all those binaries, extending Nakamura et. al's [19] argument the number of black holes will be $(10^{-3}, 10^{-4})$ times the number of SQN binaries. The geometrical arguments with the dynamics of black holes of Nakamura et. al. [19] is essentially valid in the present case. Nakamura et. al. [19] deals with BHMACHO (black hole macho), here I am dealing with SQNs, the probability distribution remains essentially the same. Thus in the universe there will be approximately ~ $10^{17\text{-}18}$ black holes originating from MACHOs.

Similarly, for the milky way out of $N_{macho}$ ~ $10^{13\text{-}14}$, $10^{9\text{-}10}$ would be black holes.

For the neutron stars, out of $10^{9\text{-}10}$ neutron stars in the universe $(10^{5\text{-}6})$ will be binary system, eventually leading $10^5 M_\odot$ quark matter [1].

How does one distinguish between black hole BHSQN from the SQNs and the SQN binaries which are not black holes?

SQNs, while forming binaries, has the radius of the binary $R_T$ less than $2R_0$, being approximately $(1.2 - 1.4) R_0$ as shown using Hawking entropy surface area relation viz eqns (1) & (3), Table 1. It is suggested that the missing significant mass, is used up as gamma ray bursts just as is expected to happen in neutron stars; the time delay, as has been noted [19], between gravitational wave and gamma ray is burst ~ 1sec. In contrast, black holes will not radiate any gamma ray at all.

If gamma rays for same reason are not emitted by coalescing SQNs, leading to a binary one may still use their observed "chirp" mass $M_{chirp}$ to distinguish them from BHSQNs [21].

***This distinction can be usefully exploited; using LIGO/VIRGO and any form of high energy gamma ray detectors, we can conclude fairly precisely whether quark like galaxy, MQG is the reality for dark matter or the only black hole galaxy is the reality for dark matter.***

This is important: if the quark like galaxy MQG idea comes through, one has found a novel form of dark matter also, one can trace back the history of the universe to primordial epoch. Second, black holes emerging from a quark like galaxy may open up an entirely an unexplored domain of cosmology. In short, using the present technology, observations of gravitational wave in conjunction with high energy gamma bursts will be the vital cursor of this scenario.

However, if BHMACHO and their binaries dominate the dark matter world, there will be no gamma ray bursts; a galaxy of black holes only as candidate of the entire dark matter is an unlikely scenario. In either cases the contribution from the newtron star core is negligible ~ $10^5 M_\odot$.

## OUTLOOK

MACHOs are candidates of Strange Quark Nuggets. In the universe there are $N_{MACHO} = 10^{23-24}$ Machos. With a mass of SQN, typically of 0.5 $M_\odot$, which seems to be the most plausible value, SQNs will populate about 12.5% of the dark matter but with a mass of $1 M_\odot$ SQNs will constitute 25% of the dark matter. This is for $N_{macho} \sim 10^{23}$, the most likely number for $N_{macho} \sim 10^{24}$ is ruled out.

With the rather precise gravitational wave interferometers like LIGO/VIRGO and others it is entirely possible with today's technology to determine (Black Hole + SQNs) like galaxy; MQG, BHSQN formed from some of the SQNs binary in conjunction with the gamma ray bursts will be a definitive cursor of this observation.

This work is a sequel to my recent work, in ref. [9]; I take this opportunity to thank all my colleagues, who have assisted me to formulate these ideas. I thank Edward Witten for his perceptive comments on supercooling, Roger Penrose for his positive encouragement and comments. Dima Kharzeev for his encouraging comments and Larry McLarren for his interest and comments. Discussions with Pijush Bhattacharya, Debasish

Mazumdar and Sibaji Raha are gratefully acknowledged. The assistance rendered by Chiranjib Barman is much appreciated.

## References

[1] E. Witten, Phys. Rev. D **30**, 272 (1984)

[2] E. Witten, Private Communication (2014).

[3] T. Boeckel and J. Schaffner-Bielich, Phys. Rev. Lett. **105**, 041301 (2010).

[4] T. Boeckel and J. Schaffner-Bielich, Phys. Rev. D **85**, 103506 (2012).

[5] N. Borghini et. al., J. Phys. G **26**, 771, (2000).

[6] P. Bhattacharjee et. al., Phys. Rev. D **48**, 4630 (1993).

[7] B. Sinha, J. Phys. Conf. Ser. **668**, 012028 (2016).

[8] P. Castorina et. al., Eur. Phys. J. C **52**, 187 (2007).

[9] B. Sinha, Phys. Rev. D **101**, 103007 (2020).

[10] S. Banerjee et. al., Mon. Not. R. Astron. Soc. **340**, 284 (2003).

[11] C. Alcock et. al., Nature (London) **365**, 621 (1993)

[12] C. Alcock et. al., Astrophys. J. **479**, 119 (1997).

[13] C. Alcock et. al., Astrophys. J. **486**, 697 (1997).

[14] S. Banerjee et. al., J. Phys. G **26**, L1 (2000).

[15] B. Sinha, Nucl. Phys. **A982**, 235 (2019).

[16] K. Jedamzik, Phys. Rep. **307**, 155 (1998).

[17] S. Hawking, Nature **248**, 30 (1974)

[18] S. Hawking, Commun. Math. Phys. **43,** 199 (1975).

[19] T. Nakamura et. al., Astrophys. J. **487**, L139 (1997).

[20] B. Sinha, to be published.

[21] K. S. Thorne, 1987, in 300 years of Gravitation ed. S. W. Hawking & W. Israel (Cambridge, Univ. Press) 330

K. S. Thorne, 1995, in Proc. Snowmass 95 summer study on Particle and Nuclear Astrophysics and Cosmology, ed. E. W. Kolb & R. Peccei, 398